\documentclass[prl,twocolumn]{revtex4-1}

\usepackage{amsmath}
\usepackage{amssymb}
\usepackage{amsthm}

\usepackage{graphicx}
\usepackage{graphics}
\usepackage{bbold,bm}
\usepackage{hyperref}
\usepackage{epsfig}
\usepackage{dcolumn}

\newcommand\vex[1]{\mathbf{#1}}

\def\sgn{\mathrm{sgn}}

\def\dd{\mathrm{d}}

\def\dbar{\hbox{$d$\kern-0.6em\raise0.3em\hbox{$-$}}\hspace{-0.5mm}}

\begin{document}

\title{Majorana edge modes of topological exciton condensate with superconductors}

\author{Babak Seradjeh}

\address{Department of Physics, Indiana University, 727 East Third Street, Bloomington, IN 47405-7105 USA}

\begin{abstract}
I study the edge states of the topological exciton condensate formed by Coulomb interaction between two parallel surfaces of a strong topological insulator. When the condensate is contacted by superconductors with a $\pi$ phase shift across the two surfaces, a pair of counter-propagating Majorana modes close the gap at the boundary. I propose a nano-structured system of topological insulators and superconductors to realize unpaired Majorana fermions. The Majorana signal can be used to detect the formation of the topological exciton condensate.  The relevant experimental signatures as well as implications for related systems are discussed.
\end{abstract}

\maketitle

%---- Introduction
\emph{Introduction.}---Topological states of matter exhibit a novel form of quantum order, leading to precise quantization of certain physical quantities~\cite{Wen04a}.  Important examples of such \emph{topologically} ordered states are provided by the hierarchy of quantum hall states~\cite{Hal83a,Hal84a,BloWen90b} and time-reversal invariant topological insulators (TIs)~\cite{HasKan10a,QiZha11a,HasMoo11a}.
%and unusual quantum numbers assigned to quasiparticles. 
The combination of topological order with the conventional order arising from broken symmetries gives rise to new degrees of freedom, such as protected chiral or helical surface states and Majorana modes, that are otherwise not realized in nature. The study of these states is crucial to our understanding of the collective properties of matter and could be important for applications ranging from high-precision metrology to quantum computation.

Recently, a topological exciton condensate (TEC) was predicted by a group including the author to exist when a thin film of strong TI is gated so that electron and hole gases separately form on opposite surfaces due to Coulomb interaction~\cite{SerMooFra09a,Ser12a}. In the exciton condensate state, the system acquires a coherent tunneling amplitude between the two surfaces even though there is no direct electronic  tunneling between them. It was shown that the topological nature of the TEC gives rise to fractionally charged vortices of the condensate. A similar system based on graphene has also beed proposed and studied~\cite{LozSok08a,MinBosSu08a,SerWebFra08a}; however, due to the higher multiplicity of Dirac cones, the exciton condensate in graphene is topologically trivial. 

\begin{figure}[t]
\begin{center}
\includegraphics[scale=0.95]{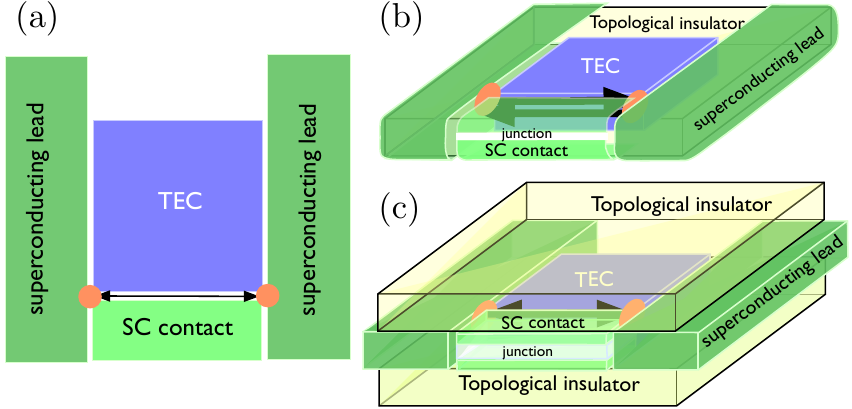}
\end{center}
\caption{Majorana fermions. The TEC is attached to superconducting leads and contacts with a phase difference $\delta\varphi$ between the top and bottom surfaces, (a) top view, and the proposed setup in (b) thin film TI and (c) sandwich structures, where a dielectric wafer is sandwiched between two TI slabs. By tuning $\delta\varphi$ in the middle contact, a pair of counter-propagating Majorana modes (arrows) form at the boundary with the TEC, resulting in two unpaired Majorana fermions (dots) at the intersection.}\label{fig:setup}
\end{figure}

In this Letter, I study the structure of edge states when the TEC is interfaced with other ordered states and propose a novel way to detect the TEC experimentally. Remarkably, I show that under certain conditions a (nonchiral) Majorana channel exists at the boundary of the TEC with a superconductor that spans the two surfaces. In the lab, it might be better to substitute the thin film with a dielectric wafer sandwiched between two TIs. This design offers greater control on the dielectric screening of the insulating spacer~\cite{ChoMoo11a}. Fig.~\ref{fig:setup} shows the schematics of the proposed setup for both a thin-film and a sandwich structure. As the superconducting phase shift across the Josephson junction is cycled through $2\pi$, the Majorana channel opens once, resulting in two unpaired Majorana modes moving along the channel in opposite directions. These Majorana modes exhibit non-Abelian fractional statistics~\cite{Iva01a} and may be detected using a suite of experimental techniques that are currently used or being developed~\cite{SasKriSeg11a,KorKirLah11a,WilBesGal12a,Kou12a}. This prediction can be exploited either to realize and manipulate unpaired Majorana fermions or, more significantly, to detect the TEC itself.

\begin{figure}[tb]
\begin{center}
\includegraphics[width=3.4in]{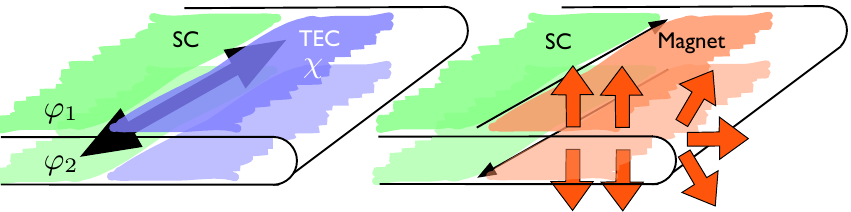}
\end{center}
\caption{Constructing the Majorana edge modes by bending a single surface, starting from (left) a domain wall in the single-layer superconductor with $\varphi_1-\varphi_2=\pi$, and (right) the boundary between superconducting and magnetic regions. }\label{fig:magnet}
\end{figure}

The existence of the Majorana channel may be understood in terms of the Majorana edge modes on an isolated surface of a TI, which are found in the following two cases. First, there is a pair of counter-propagating Majorana edge modes at a domain wall separating (proximity-induced) superconducting regions with a phase difference $\pi$~\cite{FuKan08a}. As shown in Fig.~\ref{fig:magnet}(a) by folding this domain wall across the side surface to the opposite surface these Majorana edge modes map to the ones described above. Second, there is a \emph{chiral} Majorana edge state at the boundary of a superconductor on a single surface and a magnetic region dominated by the Zeeman energy. As shown in Fig.~\ref{fig:magnet}(b), by folding the magnetic region as before, we obtain a pair of counter-propagating Majorana edge modes in each surface living at the boundary between superconducting and \emph{antiparallel} magnetic regions on the two surfaces. The mass term associated with the antiparallel magnetic order, indicated by $M_a$ in Eq.~(\ref{eq:h0}) below, anticommutes with the TEC mass term, i.e. they add in squares to give the square of the energy gap in a region with both orders. Therefore, we may adiabatically switch the TEC on and the magnetic region off without closing the bulk gap. Since this process couples the two surfaces through exciton tunneling terms, the pair of Majorana modes will survive only if protected by a symmetry. I will show that this is precisely what happens when the phase of the superconductor between the two surfaces is adjusted to $\pi$ and the symmetry is closely related to time-reversal (T) symmetry.

%----------------
\emph{Mass terms.}---The Hamiltonian describing the surface states of the system is $H=\sum_{\vex r} \psi^\dagger(\vex r) \hat{h}_0 \psi(\vex r)$ where 
%%%%
\begin{equation}\label{eq:h0}
\hat{h}_0 = \tau_z (v\boldsymbol{\sigma}\cdot\hat{\vex  p} +V) + |m| \tau_x e^{i\chi\tau_z} + M, 
\end{equation}
%%%%
and the spinor $\psi^\intercal = e^{i\pi\sigma_z/4}(\psi_{1\uparrow},\psi_{1\downarrow},\psi_{2\uparrow},\psi_{2\downarrow})$ has four components indexed by surface ($\alpha=1,2$) and spin ($\uparrow$,$\downarrow$) labels, with $v$ the Fermi velocity, $\boldsymbol{\tau}$ and $\boldsymbol{\sigma}$ Pauli matrices acting on the surface and spin index, respectively, and $\hat{\vex p}=-i(\partial_x,\partial_y)$ the momentum operator. Here, $|m|e^{i\chi}$ is the complex exciton order parameter, $V$ is a symmetric bias, and I have included the magnetic order parameters $M=M_p\sigma_z +M_a\tau_z\sigma_z$ for the parallel ($M_p$) and antiparallel ($M_a$) components along the surface normal.

To include the superconducting order parameter, we pass to the Nambu spinor $\Psi^\intercal = (\psi^\intercal, i\psi^\dagger\sigma_y )$ and the Bogoliubov--de~Gennes Hamiltonian 
%%%%
$
\hat h = \left(\begin{array}{cc} \hat h_0 & \Delta^\dagger \\ \Delta & -\sigma_y \hat h_0^\intercal \sigma_y \end{array}\right).
$
%%%%
The superconducting order parameter in each surface has the form $|\Delta_\alpha|e^{i\varphi_\alpha}$, so that 
%%%%
\begin{equation}
\Delta = |\bar{\Delta}|e^{i\varphi}\tilde\Delta^{\tau_z}e^{i\delta\varphi\tau_z/2},
\end{equation}
%%%%
where $|\bar\Delta|=\sqrt{|\Delta_1\Delta_2|}$, $\tilde\Delta=\sqrt{|\Delta_1/\Delta_2|}$, the total phase $\varphi=\frac12(\varphi_1+\varphi_2)$ and the phase shift $\delta\varphi=\varphi_1-\varphi_2$, and $\boldsymbol{\eta}$ are Pauli matrices acting on the Nambu index. I set $\tilde\Delta=1$, for now, and consider other values later. We may write the full Hamiltonian as %$\Psi\mapsto e^{i\varphi\eta_z/2}\Psi$, 
$\hat h = U^\dagger(\hat h_s + M)U$, with $U=e^{i\varphi\eta_z/2}e^{i\chi\eta_z\tau_z/2}$ and
%%%%
\begin{equation}\label{eq:hs}
\hat h_s = \eta_z\tau_z (v\boldsymbol{\sigma}\cdot\hat{\vex p} + V) + |m|\eta_z\tau_x + |\Delta|\eta_x e^{i(\delta\varphi/2-\chi)\eta_z\tau_z},
\end{equation}
%%%%
which shows only the combination $\delta\varphi-2\chi$ is physically significant.

These eight mass terms leave out two additional superconducting mass terms  $\sim \eta_x\tau_y\sigma_z$ and $\eta_y\tau_y\sigma_z$ that mediate pairing between the two surfaces. They may be  generated either by interlayer proximity effect via the superconductor or by an interlayer pairing interaction. I assume these processes are negligible.

\emph{Symmetries.}---%Hamiltonian~(\ref{eq:hs}) has a number of important symmetry properties.
The T symmetry is given by the antiunitary operator $\Theta=i\sigma_yK$ with $K$ the complex conjugation and is broken for a general value $\delta\varphi-2\chi\neq 2n\pi$, $n\in\mathbb{Z}$. Physically, these values amount to having an inplane flux. For $\delta\varphi-2\chi=(2n+1)\pi$, $\Theta$ anticommutes with the last term in Eq.~(\ref{eq:hs}). Note that $\eta_z$ also commutes with all the terms except the last term with which it anticommutes. Therefore at these special values, a new symmetry is obtained given by the antiunitary operator $\Upsilon=\eta_z\Theta$. This will be important for discussing the edge modes.

%--------------
\emph{Majorana edge modes.}---The edge modes for various phase boundaries can be understood in terms of the algebraic relations of the corresponding mass terms: if they anticommute, the gap will never close at the boundary and hence there are no gapless edge modes; if they commute, the gap closes at the boundary and a gapless edge channel opens. To see this, let us consider a boundary along the $y$ axis with $|m|-|\Delta|<0$ for $x\to\infty$ and $>0$ for $x\to-\infty$ and set $M=0$. For $V=0$, we find the edge modes
%%%%
\begin{equation}\label{eq:edge}
\Psi(p_y) = C e^{ip_y y} e^{\int^x \left[|m(r)|-|\Delta(r)\sin(\delta\varphi/2-\chi)|\right]\dd r/v}\Psi_0,
\end{equation}
%%%%
where $\tau_y\sigma_x\Psi_0=s\eta_x\sigma_x\Psi_0=-\Psi_0$, $s=\sgn[\sin(\delta\varphi/2-\chi)]$, and $C$ is a normalization factor.  This eigenvalue problem has two solutions, which can be chosen as eigenstates $\eta_z\tau_z\sigma_y\Psi^\pm_0=\pm\Psi^\pm_0$, since $\eta_z\tau_z\sigma_y$ commutes with both $\tau_y\sigma_x$ and $\eta_x\sigma_x$. Projecting to the subspace spanned by $\Psi^\pm_0$ on which Pauli matrices $\boldsymbol{\rho}$ act, the Hamiltonian takes the form 
%%%%
\begin{equation}\label{eq:hs0}
\hat h_s|_0=v \rho_z p_y + \delta\cos(\delta\varphi/2-\chi)\rho_x,
\end{equation}
%%%%
where the overlap $\delta=\int\Psi^-_0{}^\dagger(x)\eta_x\Psi^+_0(x) |\Delta(x)| \dd x$. The energy of the bound states is then found to be
%%%%
\begin{equation}\label{eq:spec}
E(p_y) = \pm\sqrt{v^2p_y^2+\delta^2\cos^2(\delta\varphi/2-\chi)}.
\end{equation}
%%%%

\begin{figure}[tb]
\begin{center}
\vspace{5mm}
\includegraphics[scale=1]{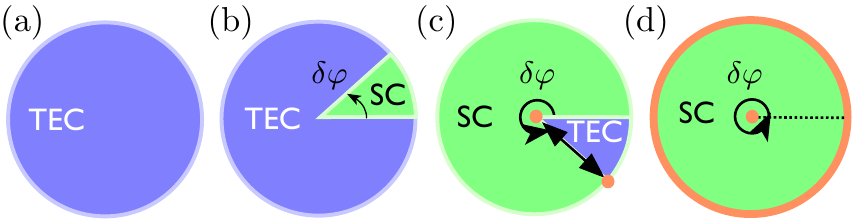}
\end{center}
\caption{Unpaired Majorana fermion. Starting from (a) a droplet of TEC, one (b) condenses a pie of the superconducting state with the phase $\delta\varphi$ winding from 0 to the pie angle. At (c) $\delta\varphi-2\chi=\pi$ and a pair of Majorana modes form along the radius. When (d) the droplet is filled with the superconductor with a $2\pi$ vortex in $\delta\varphi$ an unpaired Majorana fermion sits at the vortex core and one at the boundary. %Restricting the total phase $\varphi=0$ results in a branch cut shown by the dotted line.
}\label{fig:SC-vortex}
\end{figure}

When $\delta\varphi-2\chi=(2n+1)\pi$, the edge states form a pair of counter-propagating gapless Majorana modes along the edge. %This happens once each time $\delta\varphi$ is cycled through $2\pi$, independent of the value of $\chi$.
The degeneracy at $p_y=0$ is protected by $\Upsilon$ symmetry. The two solutions $\Psi_0^\pm$ are $\Upsilon$-partners, i.e. $\Upsilon \Psi_0^\pm = \pm \Psi_0^\mp$. Since $\Upsilon^2=-1$, Kramers theorem applies and ensures that a pair of zero-energy states survive in the presence of potentials that do not break $\Upsilon$ symmetry. Since $\Upsilon$ is fundamentally connected to $\Theta$, it seems this protection should be equally robust as the one under T symmetry. As a direct consequence, since $\Upsilon \eta_z\tau_z \Upsilon^{-1} = \eta_z\tau_z$, a non-zero $V$-term will not split the degeneracy. A $\tilde\Delta\neq 1$ introduces an additional term $\sim \eta_y$ in Eq.~(\ref{eq:hs}) for $\delta\varphi-2\chi=(2n+1)\pi$. Again, since $\eta_y=\Upsilon\eta_y\Upsilon^{-1}$, the gapless edge states persist.

%-----------------
\emph{Unpaired Majorana fermions.}---The above discussion suggests that an unpaired Majorana fermion must exist when $\delta\varphi$ contains a vortex of winding $2\pi$. Imagine starting with a circular droplet of TEC, as in Fig.~\ref{fig:SC-vortex} and slowly opening a pie with the superconducting order parameter where $\delta\varphi$ varies from $0$ up to the pie angle. As the  angle ranges from $0$ to $2\pi$, at some point along the process the condition $\delta\varphi-2\chi=\pi$ is satisfied and a channel of counter-propagating Majorana modes open along the radius, thereby localizing a Majorana fermion at the center of the droplet and another at the boundary. Finally, the pie closes onto itself, creating a vortex of winding $2\pi$ in $\delta\varphi$ and leaving a \emph{single} Majorana fermion at the vortex core and its partner at the boundary.

In Hamiltonian~(\ref{eq:hs}), since we have mapped to $\varphi=0$, the vortex in $\delta\varphi$ consists of $\pm\pi$ windings of the superconducting phase for each surface. Therefore, this symmetric description introduces branch cuts in the order parameter. A nonsingular description is obtained if $\varphi$ also contains a $\pm\pi$ winding, so that there is a full quantum vortex in one surface and none in the other. In other words, the vortex in $\delta\phi$ appears as a magnetic monopole with flux $h/2e$ penetrating one surface and then spreading between the planes. The single Majorana mode then lives in the surface containing the vortex~\cite{FuKan08a}. Our discussion shows that this Majorana bound state is stable in general when there is a vortex in $\delta\varphi$ regardless of how it is distributed between the two surfaces.

\begin{figure}[tb]
\begin{center}
\includegraphics[scale=1]{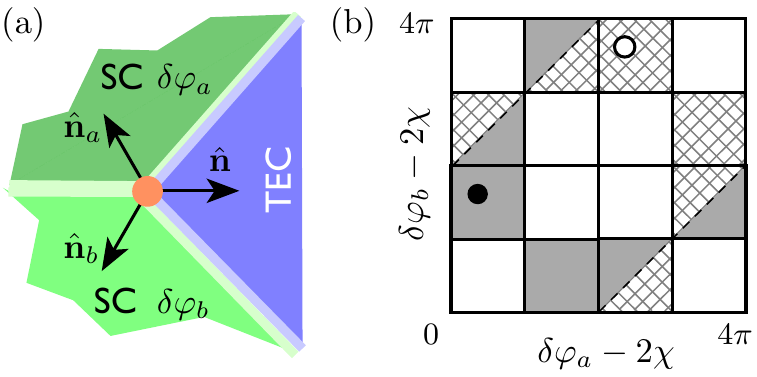}
\end{center}
\caption{(a) Unpaired Majorana fermion (dot) at the junction of two superconducting contacts and the exciton condensate (top view). (b) The topological phase diagram where regions with orthogonal unpaired Majorana fermions are shaded and hashed. The solid and open dots mark the values at which an analytical solution for the Majorana bound state is given in Eq.~(\ref{eq:MBS}).}\label{fig:junction}
\end{figure}

We can check this claim explicitly, e.g. for the junction geometry in Fig.~\ref{fig:junction}(a) where the exciton order parameter is $|m|\tau_xe^{i\chi\tau_z}$ for $x>0, |y|<\sqrt3|x|$ and zero otherwise, and the superconducting order parameter is $|\Delta|e^{i\delta\varphi\tau_z/2}$ for $x<0, |y|>\sqrt3|x|$ and zero otherwise with $\delta\varphi=\delta\varphi_a$ for $y>0$, and $\delta\varphi_b$ for $y<0$. Assuming the junctions have vanishing width and setting $V=0$, we see from Eq.~(\ref{eq:hs0}) that a bound state exists when there is a kink in the $\rho_x$ mass term, i.e. when $\cos(\delta\varphi_a/2-\chi)\cos(\delta\varphi_b/2-\chi)<0$. For $\delta\varphi_a-2\chi=\pi/3$ and $\delta\varphi_b-2\chi=5\pi/3$, shown by the dot in Fig.~\ref{fig:junction}(b), and $|\Delta|=|m|$ an analytical solution for the Majorana bound state, $\hat h_s\Gamma=0$, may be found as
%%%%
\begin{equation}\label{eq:MBS}
\Gamma = C e^{-|m|\hat{\vex n} \cdot \vex r/ v} \gamma,
\end{equation}
%%%%
where $\gamma = \frac1{\sqrt2}(\Psi_0^+-ic\Psi_0^-)$ and $\hat{\vex n}$ is a unit vector bisecting each region as shown in Fig.~\ref{fig:junction}(a). There is only one solution and the sign $c=\sgn[\cos(\delta\varphi_a/2-\chi)]=+1$ is chosen by the orientation of the superconducting regions and the exciton condensate. Finally, non-zero $V$ will not remove this zero-energy state since the spectrum is particle-hole symmetric, $\Omega\hat h_s \Omega^{-1} = -\hat h_s$ with $\Omega=\eta_y\sigma_yK$. 

The bound state is stable as $\delta\varphi_a$ and $\delta\varphi_b$ change so long as the gap along the boundaries does not close. The Hamiltonian~(\ref{eq:hs}) is $4\pi$-periodic in $\delta\varphi-2\chi$. This is  a manifestation of the fractional Josephson effect due to Majorana modes~\cite{Kit01a,FuKan09a}. Note that when $|\delta\varphi_a-\delta\varphi_b|=2\pi$ there will be a domain wall along the boundary of the two superconductors. This leads to \emph{two} nonchiral Majorana channels along that edge. While this does not change the parity of the unpaired Majorana fermions, it does change the superposition of $\Psi^{\pm}_0$ in $\gamma$ to an orthogonal combination. For example, $c$ changes from $+1$ to $-1$ when $\delta\varphi_a-2\chi=7\pi/3$ and $\delta\varphi_a-2\chi=11\pi/3$, shown by an open dot in Fig.~\ref{fig:junction}(b). Combining these facts, we find the topological phase diagram shown in Fig.~\ref{fig:junction}(b).

%-----------------
\emph{Discussion.}---Although expected from our general arguments, it is still remarkable that we find a \emph{single} Majorana bound state for an $8\times 8$ underlying Hamiltonian, $\hat h_s$. This is the same dimensionality as the low-energy Hamiltonian of superconducting state of spinless fermions on a honeycomb lattice (such as graphene), where the role of spin and exciton order parameter is played by the sublattice pseudospin and the Kekul\'e distortion, respectively. Previous studies found an even number of Majorana fermions in such a system~\cite{GhaWil07a,BerLe-09a}. In our case, the superconducting order parameter with phase difference $\delta\varphi$ across the two surfaces translates to a superconducting state with intra-valley pairing and phase difference $\delta\varphi$ across the two valleys. The pairs in this superconducting state of graphene have finite momentum. A related system is the new family of iron pnictide high-temperature superconductors where there is evidence for $s^\pm$ pairing. Indeed, a recent study~\cite{DenVioOrt12a} reached similar conclusions in a lattice model with an $8\times 8$ Hamiltonian. However, it seems difficult practically to engineer a vortex configuration in these systems, since the phase shifts that need to be tuned occur in momentum space.

These findings can be utilized in two ways. First, they can be used to create and manipulate Majorana fermions. The protocols for doing so are similar to those proposed by Fu and Kane~\cite{FuKan08a}. However, in regard of the increased complexity of the device designs, this is not a particularly advantageous application. Second, and more significantly, the detection of the Majorana modes can be used as a proxy for the TEC. This opens the way to new techniques~\cite{SasKriSeg11a,KorKirLah11a,WilBesGal12a,Kou12a} to signal the formation of the TEC, which are not possible for other schemes of creating an exciton condensate, such as quantum Hall bilayers, graphene, or semiconductor quantum wells. I will briefly outline the procedure, discussing separately the case of a thin film and the sandwich structures of Fig.~\ref{fig:setup}. In all these cases, controlling the phase shift $\delta\varphi-2\chi$ is vital. This can be achieved by, say, an in-plane field, along the intended Majorana channel.

In a thin film, when the two opposite surfaces are so close that there is direct electronic tunneling between them, the bulk TI is lost. There will be no Majorana signal in this case as its existence relies on there being helical surface states of the bulk TI. (Note that in the thin film the superconducting contacts must be placed outside the tunneling region.) The TEC is expected to form when the two surfaces are far enough so that there is no direct tunneling but close enough so that the Coulomb interaction between the surfaces is not negligible. In this case, the Majorana signal can be used to infer the existence of interaction-mediated tunneling between the surfaces.

In the sandwich structure, when there is direct tunneling between the two surfaces, bulk TI behavior is expected to permeate across the spacer. There is a Majorana signal in this situation. Nevertheless, it is possible to differentiate this single-particle tunneling from the interaction-mediated tunneling of TEC by studying the temperature dependence of the Majorana signal. The important point is that the bulk TI gap, which is $\sim300$~meV in Bi$_2$Se$_3$, is much larger than the typical TEC gap, which is expected to be $\sim0.1$~meV. Therefore, a Majorana signal caused by the single-particle tunneling would be unusually robust as the temperature is increased beyond the TEC gap. It should, however, be pointed out that it is unlikely that the thickness necessary to allow separate superconducting contacts inside the spacer will be in this range.

I acknowledge early discussions with E. Fradkin, M. Gilbert, and T.~L. Hughes. I thank H. Fertig and S. Vishveshwara for useful comments and G. Ortiz for the same and also for bringing Ref.~\onlinecite{DenVioOrt12a} to my attention. This research was supported by the College of Arts and Sciences at Indiana University, Bloomington.

\emph{Note added.}---In the final stages of this project, a preprint by Meng, Vishveshwara and Hughes~\cite{MenVisHug12a} appeared where similar mass terms have been considered. However, they focus on the boundary between a magnetic domain and a tunneling region.

%\begin{widetext}
%TEXT
%\end{widetext}

\vspace{-5mm}

%-----------------

\end{document}